\documentclass
[preprint,byrevtex,showpacs,10pt,letterpaper,twocolumn,final,prl]{revtex4}%
\usepackage{amsfonts}
\usepackage{amsmath}
\usepackage{amssymb}
\usepackage[dvips]{graphicx}%
\begin{document}
\title{Compositional disorder and tranport peculiarities in the amorphous indium-oxides}
\author{U. Givan$^{1,2}$ and Z. Ovadyahu$^{1}$}
\affiliation{$^{1}$Racah Institute of Physics, The Hebrew University, Jerusalem, 91904,
Israel,$~^{2}$Max Planck Institute of Microstructure Physics, Weinberg 2,
06120, Halle, Germany}

\begin{abstract}
We present results of the disorder-induced metal-insulator-transition (MIT) in
three-dimensional amorphous indium-oxide films. The amorphous version studied
here differs from the one reported earlier [Phys. Rev. B \textbf{46}, 10917
(1992)] in that it has a much lower carrier concentration. As a measure of the
static disorder we use the dimensionless parameter $k_{F}\ell$. Thermal
annealing is employed as the experimental handle to tune the disorder. On the
metallic side of the transition, the low temperature transport exhibits
weak-localization and electron-electron correlation effects characteristic of
disordered electronic systems. These include a fractional power-law
conductivity versus temperature behavior anticipated to occur at the critical
regime of the transition. The MIT occurs at a $k_{F}\ell\approx$0.3 for both
versions of the amorphous material. However, in contrast with the results
obtained on the electron-rich version of this system, no sign of
superconductivity is seen down to $\approx$0.3K even for the most metallic
sample used in the current study. This demonstrates that using $k_{F}\ell$ as
a disorder parameter for the superconductor-insulator-transition (SIT) is an
ill defined procedure. A microstructural study of the films, employing high
resolution chemical analysis, gives evidence for spatial fluctuations of the
stoichiometry. This brings to light that, while the films are amorphous and
show excellent uniformity in transport measurements of macroscopic samples,
they contain compositional fluctuations that extend over mesoscopic scales.
These, in turn, reflect prominent variations of carrier concentrations thus
introducing an unusual type of disorder. It is argued that this
\textit{compositional} disorder may be the reason for the apparent violation
of the Ioffe-Regel criterion in the two versions of the amorphous
indium-oxide. However, more dramatic effects due to this disorder are expected
when superconductivity sets in, which are in fact consistent with the
prominent transport anomalies observed in the electron-rich version of
indium-oxide. The relevance of compositional disorder (or other agents that
are effective in spatially modulating the BCS potential) to other systems near
their SIT is discussed.

\end{abstract}
\maketitle

\section{Introduction}

The metal-insulator and the superconductor-insulator transitions (MIT and SIT
respectively) are two major representatives of quantum phase transitions that
have been extensively studied over the last decades. Disorder driven MIT
\cite{1} could be affected by a number of means; doping, pressure, stress,
etc. \cite{2,3,4}. A unique method to drive the system from the insulating to
the metallic (or superconducting) state is feasible in some metallic glasses
prepared from the gaseous phase. Amorphous films quench-condensed onto
substrates by vacuum deposition usually contain micro-voids and have a lower
mass-density than the respective equilibrium material \cite{5} and thus, on
average, smaller interatomic overlap. Thermal treatment at pre-crystallization
temperatures may then be used to reduce the free volume created by the
micro-voids \cite{5}. In this method one controls the inter-atomic overlap in
a similar vein as in employing hydrostatic pressure on a solid. The main
difference is that thermal annealing inevitably causes an irreversible volume change.

Thermal annealing has been employed to change the resistivity of amorphous
indium-oxide films while monitoring their optical properties \cite{6}.
Typically, 3-5 orders of magnitude in room temperature resistivity could be
obtained accompanying a $\approx$3\% reduction in sample thickness (measured
by x-ray interferometry \cite{6}), which is a much larger volume change than
is commonly achievable by hydrostatic pressure. Using this technique, the SIT
in a three-dimensional indium-oxide films was mapped by measuring the low
temperature conductivity as function of disorder \cite{7}. The amorphous
indium-oxide had electron density of $\simeq$10$^{\text{21}}$cm$^{\text{-3}}$
and the samples static disorder was characterized by $k_{F}\ell$ (defined in
section II). In this version of indium-oxide, superconductivity survived in
samples with $k_{F}\ell\gtrsim$0.24, and, somewhat paradoxically (but
reconcilable by an inhomogeneous scenario discussed in this paper), insulating
behavior sets in at smaller disorder, for $k_{F}\ell\lesssim$0.3. The
observation that both limits are below the Ioffe-Regel criterion motivated us
to conduct the present study employing the same technique of tuning disorder
and using the same material but with a much smaller carrier concentration. The
idea was to see whether the problem is related to the existence of
superconducting or to electron-electron (\textit{e-e)} correlations effects;
in either case one expects the results to depend on carrier concentration.

In this work we map the MIT in three dimensional amorphous indium-oxide films
with a carrier concentration that is 2-3 orders of magnitude lower than the
material used in \cite{7}. No sign of superconductivity is found down to
$T\approx$0.3K, and normal transport properties are observed for both the
insulating and metallic regimes. However, the critical disorder, separating
the metallic from the insulating regime, still appears to be given by
$k_{F}\ell\approx$0.3 as in the electron-rich phase. By comparison, the MIT in
the previously studied \cite{8} crystalline version of indium-oxide occurs at
$k_{F}\ell\approx$0.75, which led us to examine the structural differences
between the two phases that might account for the different $k_{F}\ell$ at the transition.

We present data on the microstructure of both the electron-rich and the low
carrier-concentration versions of amorphous indium oxide using high resolution
microscopy and energy-dispersive local probe. Chemical analysis reveals that
the indium-oxygen ratio fluctuates across the sample on mesoscopic scales.
Combining this information with Rutherford backscattering and Hall effect
measurements suggests that large spatial fluctuations of carrier concentration
exist in these amorphous samples. The specifics of this inhomogeneity is
argued to be a natural cause for a variety of transport anomalies in these as
well as in other systems where superconductivity plays a role at
experimentally accessible temperatures.

\section{Samples preparation and measurements techniques}

The In$_{x}$O films used here were e-gun evaporated on room-temperature
microscope-slides using 99.999\% pure In$_{2}$O$_{3}$ sputtering target
pieces. Deposition was carried out at the ambience of (2-5)$\cdot
$10$^{\text{-4}}$ Torr oxygen pressure maintained by leaking 99.9\% pure
O$_{2}$ through a needle valve into the vacuum chamber (base pressure $\simeq
$10$^{\text{-6}}$ Torr). Rates of deposition used for the samples reported
here were typically 0.18-0.8~\AA /s. For this range of
rate-to-oxygen-pressure, the In$_{x}$O samples had carrier-concentration n in
the range (5-13)$\cdot$10$^{\text{18}}$cm$^{\text{-1}}$ measured by
Hall-Effect at room temperature. Electron-rich In$_{x}$O samples used in the
study (with carrier-concentration in the 10$^{\text{21}}$cm$^{\text{-3}}$
range) were produced with the conditions described earlier \cite{7}. The films
thickness in this study was 900-1500~\AA ,~which makes them effectively
three-dimensional (3D) down to the lowest temperature in our experiments
($\cong$0.3K) for most of the samples used in this work.

The as-deposited samples had extremely high resistivity $\rho$ of the order of
10$^{\text{3}}$-10$^{\text{6}}\Omega$cm. These were barely measurable even at
room temperature. To carry out the low temperatures studies, the samples
$\rho$ had to be reduced by several orders of magnitude. This was achieved by
thermal annealing. A comprehensive description of the annealing process and
the ensuing changes in the material microstructure are described elsewhere
\cite{6,9}. For completeness, we give here the basic protocol that was used in
this study. Following deposition and initial conductance measurement that, in
many cases required the use of electrometer (Keithley 617), the sample was
attached to a hot-stage at a constant temperature T$_{a}$, initially 5-10
degrees above room temperature. The resistance R of the sample was observed to
slowly decrease over time. T$_{a}$was raised by few degrees whenever
$\Delta\rho/\rho$ over 24 hours was less than 1\% (and the value of the
resistance was still higher than desired). To obtain a sample with $\rho$ that
was useful for the measurements reported here usually took 20-38 thermal
cycles. The annealing temperature T$_{a}$ was limited to $\approx$370~K to
minimize the risk of crystallization. The amorphicity of the samples during
the annealing process was monitored by checking the diffraction pattern of a
controlled specimen prepared on a carbon-coated copper grid. The control
specimen was deposited simultaneously with the sample used for the transport
measurement but its thickness was limited to 200-300~\AA ~to facilitate high
resolution microscopy. Electron-diffraction micrograph of the as-deposited
material is shown in Fig.~1a along with a micrograph taken for the same grid
after a prolong period of thermal annealing resulting in resistance change of
more than three orders of magnitude. Note the characteristic broad rings of
the amorphous phase, and in particular the absence of any diffraction ring of
either crystalline indium-oxide or metallic indium.%
%TCIMACRO{\FRAME{ftbpFU}{3.5181in}{5.6126in}{0pt}{\Qcb{Electron diffraction
%patterns of (a) as-prepared In$_{x}$O film and (b) the same film after thermal
%annealing at temperatures between 330 to 370K for 34 days. The as-deposited
%sample used for transport had room-temperature two-terminal resistance larger
%than 120M$\Omega$ and it dropped to 70k$\Omega$ after annealing.}}%
%{}{fig_1.eps}{\special{ language "Scientific Word";  type "GRAPHIC";
%maintain-aspect-ratio TRUE;  display "PICT";  valid_file "F";
%width 3.5181in;  height 5.6126in;  depth 0pt;  original-width 6.9401in;
%original-height 8.348in;  cropleft "0";  croptop "1";  cropright "1";
%cropbottom "0";  filename 'fig_1.eps';file-properties "XNPEU";}}}%
%BeginExpansion
\begin{figure}
[ptb]
\begin{center}
\includegraphics[
height=5.6126in,
width=3.5181in
]%
{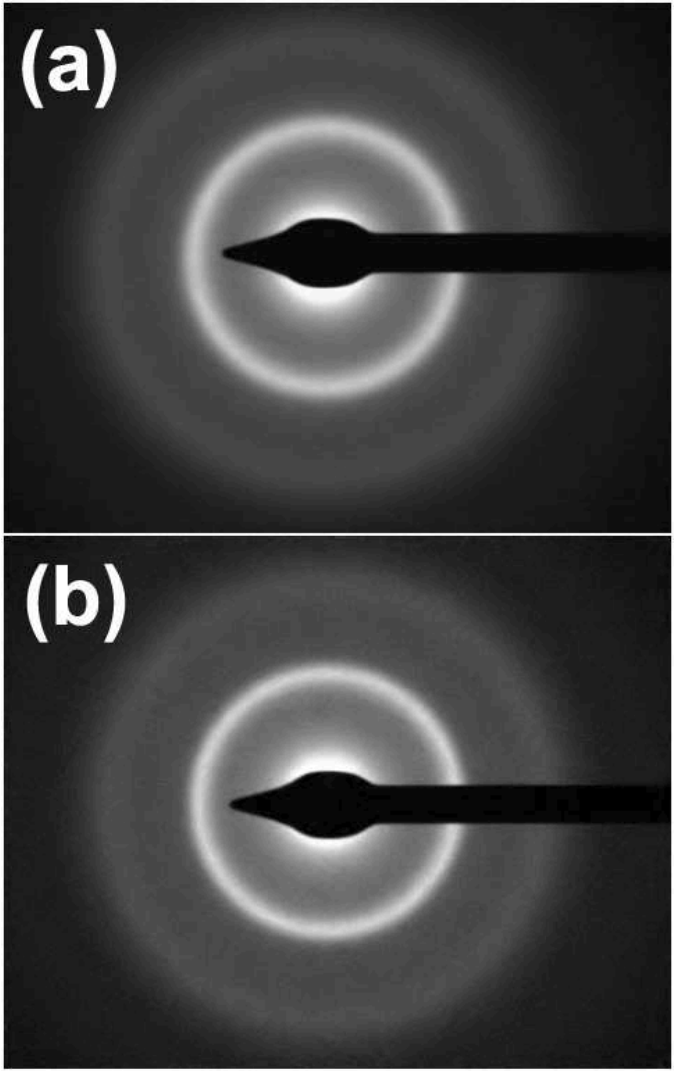}%
\caption{Electron diffraction patterns of (a) as-prepared In$_{x}$O film and
(b) the same film after thermal annealing at temperatures between 330 to 370K
for 34 days. The as-deposited sample used for transport had room-temperature
two-terminal resistance larger than 120M$\Omega$ and it dropped to 70k$\Omega$
after annealing.}%
\end{center}
\end{figure}
%EndExpansion

Conductance and Hall Effect measurement were carried out on samples that were
patterned in a 6-probe configuration using stainless-steel masks during
deposition. We used a standard Hall-bar geometry with the active channel being
a strip of either 0.6 mm or 1 mm wide, and 10 mm long. The two pairs of
voltage probes (that doubled as Hall-probes), were spaced 3 mm from one
another along the strip. This arrangement allowed us to assess the large scale
uniformity of the samples, both in terms the longitudinal conductance and the
Hall effect. Excellent uniformity was found on these scales; resistivities of
samples separated by 1~mm along the strip were identical to within $\pm$2\%
for all the samples used in the study. It should be noted however that
microstructural studies performed in the course of the study revealed
prominent inhomogeneities on mesoscopic scales (10$^{\text{2}}$-10$^{\text{3}%
}$~\AA ). The effects of these medium-scale irregularities on electronic
transport are discussed in the next section.

Most of the conductance versus temperature $\sigma(T)$ measurements were done
in the 1.3$\leq$T$\leq$15K range by a four terminal dc technique using
Keithley K220 current-source while monitoring the voltage with Keithley K2000.
For lower temperatures measurements (down to $\approx$280 mK), we used a four
terminal ac technique with the lock-in PAR124A. In all cases care was taken to
maintain linear-response conditions by keeping the voltage across the sample
low enough. This was verified by measuring the current-voltage characteristics
at the lowest temperature of the experiment.

As in previous studies \cite{6,7,8}, we use in this work $k_{F}\ell$%
=(3$\pi^{\text{2}}$)$^{\text{2/3}}\frac{\hbar\sigma_{\text{RT}}}{e^{\text{2}%
}n^{\text{1/3}}}$ as a dimensionless measure of the material disorder. This is
based on free-electron expressions using the measured room-temperature
conductivity $\sigma_{\text{RT}}$ and the carrier-concentration n, obtained
from Hall-Effect measurements, as input parameters. More details of
preparation and characterization of In$_{x}$O samples are given elsewhere
\cite{6}.

\section{Results and discussion}

One of the main goals of the research was to identify the critical $k_{F}\ell$
at which the metal-to-insulator transition occurs in the low-n version of
In$_{x}$O. This was accomplished by measuring $\sigma(T)$ over a certain range
of temperatures, then extrapolate the data~to $\sigma_{0}\equiv\sigma
$($T\rightarrow$0) to determine whether the system is insulating ($\sigma
_{0}\leq$0) or metallic ($\sigma_{0}$%
%TCIMACRO{\TEXTsymbol{>}}%
%BeginExpansion
$>$%
%EndExpansion
0$)$. Four different deposition batches were used in this phase of the
research. In each of these, thermal annealing was used, in effect, generating
samples with different $k_{F}\ell$ from the same physical specimen. A specific
series of $\sigma(T)$ for samples labeled by their $k_{F}\ell$ values is shown
in Fig.~2. Note that the $\sigma(T)$ plots for samples with $k_{F}\ell\gtrsim
$0.4 are nearly parallel to each other, a feature commonly found in many
systems on the metallic side of the metal-insulator transition. The usual
practice adopted in these cases is to extract the value of $\sigma_{0}$ by
fitting the measured $\sigma(T)$ for a given sample to:%
\begin{subequations}
\begin{equation}
\sigma(T)=\sigma_{0}+A\cdot T^{\frac{1}{2}}\tag{1}\label{eq1}%
\end{equation}%
%TCIMACRO{\FRAME{ftbpFU}{3.5405in}{2.8807in}{0pt}{\Qcb{Conductivity versus
%temperature plots for several In$_{x}$O films labeled by their $k_{F}\ell$
%values. Dashed lines are fits to Eq.1 (see text).}}{}{fig_2.eps}%
%{\special{ language "Scientific Word";  type "GRAPHIC";  display "PICT";
%valid_file "F";  width 3.5405in;  height 2.8807in;  depth 0pt;
%original-width 10.6138in;  original-height 8.1327in;  cropleft "0";
%filename 'fig_2.eps';file-properties "XNPEU";}}}%
%BeginExpansion
\begin{figure}
[ptb]
\begin{center}
\includegraphics[
height=2.8807in,
width=3.5405in
]%
{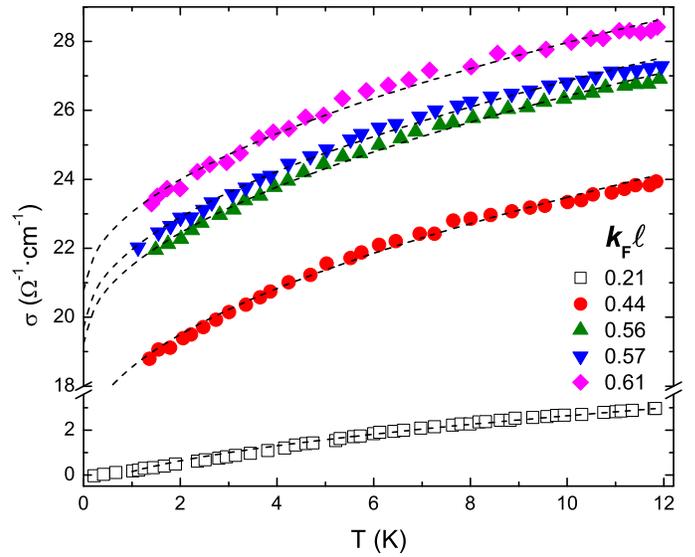}%
\caption{Conductivity versus temperature plots for several In$_{x}$O films
labeled by their $k_{F}\ell$ values. Dashed lines are fits to Eq.1 (see
text).}%
\end{center}
\end{figure}
%EndExpansion
that presumably describes transport corrections to the conductivity due to
either weak-localization or \textit{e-e} interactions effects \cite{1}. Eq.~1
offers a reasonably good fit to the data of samples having $k_{F}\ell\gtrsim
$0.4 (Fig.~2).

However, the temperature dependence of the conductance for samples that are in
the immediate vicinity of the transition suggests that another $\sigma(T)$ law
might be relevant, and that should perhaps be taken into account in
extrapolating $\sigma(T)$ to zero temperature. So, before discussing the
$\sigma_{0}$ versus $k_{F}\ell$ results we digress now to examine the data,
shown in Fig.~3, for samples in the critical regime of the transition.%
%TCIMACRO{\FRAME{ftbpFU}{3.5129in}{2.7466in}{0pt}{\Qcb{Conductivity versus
%temperature plots for In$_{x}$O films. (a) Samples with disorder nearest the
%MIT (b) The insulating sample in (a) plotted differently to show it obeys the
%Mott hopping law of a three-dimensional sample; $\sigma(T)\propto\exp
%[-(T_{M}/T)^{1/4}].$The associated localization length of this sample $\xi$
%was estimated by the Mott's formula \cite{5} using the experimental value for
%$T_{M}.$(c) Illustrating the broad temperature range over which the sample
%with $k_{F}\ell$=0.32 (which is closest to the transition), exhibits the
%power-law behavior $\sigma(T)\propto T^{1/3}.$}}{}{fig_3.eps}%
%maintain-aspect-ratio TRUE;  display "PICT";  valid_file "F";
%width 3.5129in;  height 2.7466in;  depth 0pt;  original-width 10.1953in;
%original-height 7.952in;  cropleft "0";  croptop "1";  cropright "1";
%cropbottom "0";  filename 'fig_3.eps';file-properties "XNPEU";}}}%
%BeginExpansion
\begin{figure}
[ptb]
\begin{center}
\includegraphics[
height=2.7466in,
width=3.5129in
]%
{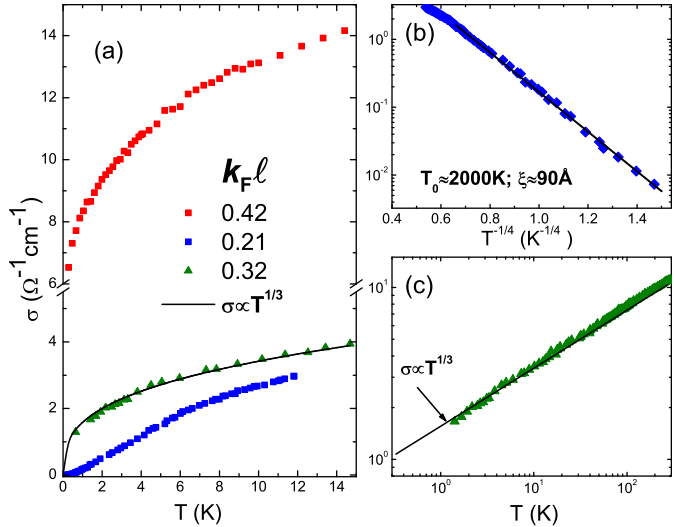}%
\caption{Conductivity versus temperature plots for In$_{x}$O films. (a)
Samples with disorder nearest the MIT (b) The insulating sample in (a) plotted
differently to show it obeys the Mott hopping law of a three-dimensional
sample; $\sigma(T)\propto\exp[-(T_{M}/T)^{1/4}].$The associated localization
length of this sample $\xi$ was estimated by the Mott's formula \cite{5} using
the experimental value for $T_{M}.$(c) Illustrating the broad temperature
range over which the sample with $k_{F}\ell$=0.32 (which is closest to the
transition), exhibits the power-law behavior $\sigma(T)\propto T^{1/3}.$}%
\end{center}
\end{figure}
%EndExpansion

The figure illustrates how the form of $\sigma(T)$ changes when the system
goes from metallic to insulating behavior (Fig. 3a). The outstanding case is
the middle curve that belongs to a sample that is \textit{just} insulating; it
fits well a power-law dependence $\sigma(T)\propto T^{\frac{1}{3}}$ over a
wide range of temperatures as illustrated in Fig.~3c. A temperature dependence
of this form is consistent with the interaction effect proposed by Larkin and
Khmel'nitskii (LK) \cite{10}. These authors observed that at the transition
the contribution of Coulomb interactions to the conductance may be expressed
by:%
\end{subequations}
\begin{equation}
\sigma(T)\sim\frac{e^{\text{2}}k_{F}}{\hbar}\left(  \frac{T}{T_{F}}\right)
^{\frac{1}{\eta}} \label{eq2}%
\end{equation}
where $\eta$ is the exponent that describes the spatial dependence of the
Coulomb potential $e\phi(r)\propto r^{-\eta}$ (with~1%
%TCIMACRO{\TEXTsymbol{<}}%
%BeginExpansion
$<$%
%EndExpansion
$\eta$%
%TCIMACRO{\TEXTsymbol{<}}%
%BeginExpansion
$<$%
%EndExpansion
3). Note that this implies the lack of metallic screening at this range of
disorder, heralding the approach of the dielectric phase.

An alternative to the LK mechanism, offered by Imry \cite{11}, has been
invoked to account for a $\Delta\sigma(T)\propto T^{\frac{1}{3}}$ component
observed in three-dimensional In$_{2}$O$_{3-x}$ samples near the transition
\cite{12}. The Imry mechanism takes account of the scale dependent diffusion
that is a property of the critical regime. To obtain the 1/3 exponent, it also
assumes a linear-with-temperature electron inelastic-rate $\tau_{in}^{-1}$. A
scattering rate $\tau_{in}^{-1}\propto T$ is consistent with the results of
our magneto-conductance measurements in the range 4K%
%TCIMACRO{\TEXTsymbol{<}}%
%BeginExpansion
$<$%
%EndExpansion
$T$%
%TCIMACRO{\TEXTsymbol{<}}%
%BeginExpansion
$<$%
%EndExpansion
77K discussed later (Fig.~9).

The conductance component associated with the Imry and the
Larkin-Khmel'nitskii mechanism is proportional to $\frac{e^{\text{2}}}{\hbar
L_{in}}$ and $\frac{e^{\text{2}}}{\hbar L_{T}}$ respectively. Here $L_{in}%
$=$\sqrt{D\tau_{in}}$ is the inelastic diffusion-length ($D$ is the diffusion
constant that in the critical regime is scale-dependent), and $L_{T}$%
=$k_{F}^{-1}(\frac{T}{T_{F}})^{\frac{1}{\eta}}$ is the LK interaction-length.
The prefactors associated with $\Delta\sigma$ of these mechanisms are not
presently known so we cannot determine their relative contribution. However,
for either mechanism, the $T^{\frac{1}{3}}$ term should be observed once
$L\ll\xi_{c}$ where $\xi_{c}$ is the correlation length \cite{12} and $L$ is
the relevant scale for the measurement at hand. Therefore, over some
temperature range, $\sigma(T)$ of metallic samples that are sufficiently close
to the transition may be describable by:%
\begin{equation}
\sigma(T)\sim\sigma_{0}+A\cdot T^{\frac{1}{3}} \label{eq3}%
\end{equation}
where $\sigma_{0}\propto\frac{e^{\text{2}}}{\hbar\xi_{c}}$%
%TCIMACRO{\TEXTsymbol{>}}%
%BeginExpansion
$>$%
%EndExpansion
$0$ and $A$ is the sum of the contributions of the Imry and LK mechanisms.

As may be expected from the similarity between the two functions, the
$\sigma(T)$ of the (metallic) samples can be fitted to either Eq.~1 and Eq.~3
almost equally well except near the transition where Eq.~3 is a better fit.
For the sample with $k_{F}\ell$=0.42 (top curve in Fig.~3a), the best fit to
Eq.~3 yields a chi-square test value of $\chi^{2}\simeq$ 0.013 as compared
with $\chi^{2}\simeq$0.08 for the best fit to Eq.$~$1. In general, fitting
$\sigma(T)$ to Eq.~1 typically gave higher value for the zero-temperature
conductivity (the $k_{F}\ell$=0.42 sample data gave $\sigma_{0}\approx
$6.3$\Omega^{\text{-1}}$cm$^{\text{-1}}$ versus $\sigma_{0}\approx$%
4.1$\Omega^{\text{-1}}$cm$^{\text{-1}}$ using Eq.~1 and Eq.~3 respectively, a
$\approx$50\% difference). The difference in $\sigma_{0}$ between the two
fitting possibilities however becomes less important for larger $k_{F}\ell$,
as can be seen in Fig.~4a. For comparison, figure 4a also includes the results
of $\sigma_{0}$ versus $k_{F}\ell$ for In$_{2}$O$_{3-x}$, the crystalline
version of 3D indium-oxide.

It is noteworthy that metallic samples with $\sigma_{0}\ll\sigma_{\min}%
$=$\frac{e^{\text{2}}}{3\pi^{\text{2}}\hbar}k_{F}$ are obtained in both
systems ($k_{F}$, calculated by the free-electron expression, is $\simeq
$8$\cdot$10$^{\text{6}}$cm$^{\text{-1}}$ and $\simeq$12$\cdot$10$^{\text{6}}%
$cm$^{\text{-1}}$ for the In$_{x}$O and In$_{2}$O$_{3-x}$ respectively).
Quantum and \textit{e-e} interaction effects are therefore quite prominent in
the low temperature transport properties of these samples.

Data delineating the critical $k_{F}\ell$ for the transition in the high-n
version of In$_{x}$O are shown in Fig. 4b. This is based on the dependence of
the activation energy $T_{0}$ on $k_{F}\ell$ for samples that are on the
insulating side of the transition (but close to it) \cite{13}. These samples
exhibit \cite{7} a peculiar $\sigma(T)$ law that empirically has been fitted
to simple activation: $\sigma(T)\propto$exp$(-T_{0}/T)$. The critical
$k_{F}\ell$ may be determined in this case by extrapolating $T_{0}(k_{F}\ell)$
to zero \cite{14}.%

%TCIMACRO{\FRAME{ftbpFU}{3.5129in}{2.9603in}{0pt}{\Qcb{(a) The extrapolated
%values for the zero-temperature conductance as function of the disorder
%parameter $k_{F}\ell$. Data shown for the In$_{x}$O samples are based on
%extrapolation using Eq.1 (full diamonds) and Eq.3 (empty diamonds)
%respectively. The data for the crystalline version (circles) are shown for
%comparison (taken from ref.8). (b) The activation energy $T_{0}$ of
%electron-rich In$_{x}$O samples as function of $k_{F}\ell$ from which the
%critical disorder is estimated by extrapolation along the dashed line (see
%text for details). The dotted line depicts the qualitative dependence of
%$T_{c}$ on $k_{F}\ell$ (based on reference \cite{7} )}}{}{fig_4.eps}%
%maintain-aspect-ratio TRUE;  display "PICT";  valid_file "F";
%width 3.5129in;  height 2.9603in;  depth 0pt;  original-width 9.5008in;
%original-height 7.9952in;  cropleft "0";  croptop "1";  cropright "1";
%cropbottom "0";  filename 'fig_4.eps';file-properties "XNPEU";}}}%
%BeginExpansion
\begin{figure}
[ptb]
\begin{center}
\includegraphics[
height=2.9603in,
width=3.5129in
]%
{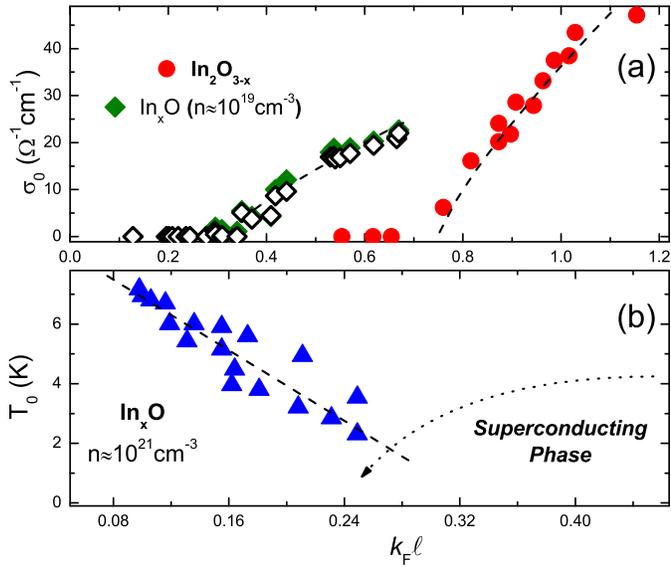}%
\caption{(a) The extrapolated values for the zero-temperature conductance as
function of the disorder parameter $k_{F}\ell$. Data shown for the In$_{x}$O
samples are based on extrapolation using Eq.1 (full diamonds) and Eq.3 (empty
diamonds) respectively. The data for the crystalline version (circles) are
shown for comparison (taken from ref.8). (b) The activation energy $T_{0}$ of
electron-rich In$_{x}$O samples as function of $k_{F}\ell$ from which the
critical disorder is estimated by extrapolation along the dashed line (see
text for details). The dotted line depicts the qualitative dependence of
$T_{c}$ on $k_{F}\ell$ (based on reference \cite{7} )}%
\end{center}
\end{figure}
%EndExpansion

Comparing the data for the three different versions of the material there are
two issues that require elucidation; The difference between the various phases
in terms of exhibiting superconductivity at the experimentally accessible
range, and the difference in the value of $k_{F}\ell$ at which the
metal-insulator transition occurs.

Note that electron-rich In$_{x}$O samples with $k_{F}\ell\gtrsim$0.3 exhibit
superconductivity for $T\lesssim$3K (Fig.~4b) while the low-n In$_{x}$O
version shows no sign of superconductivity down to $\approx$0.3K even for
samples with $k_{F}\ell$ as high as 0.68 (Fig.~5). It seems plausible that the
difference between the high-n In$_{x}$O, and the low-n version, is the large
disparity in their carrier-concentration. Likewise, the low
carrier-concentration of In$_{2}$O$_{3-x}$ (n$\approx$5$\cdot$10$^{\text{19}}%
$cm$^{\text{-3}}$) is presumably the main reason for the absence of
superconductivity in this system \cite{15}. A difference of 2-3 orders of
magnitude in carrier concentration is large enough to push down the
superconducting transition temperature $T_{c}$ well below the experimental
range; a mere factor of 3 in the BCS potential suffices to shift $T_{c}$ from
3-4K to less than 10mK.%

%TCIMACRO{\FRAME{ftbpFU}{3.5405in}{2.9482in}{0pt}{\Qcb{Conductivity as function
%of temperature plots for two In$_{x}$O films; (a) Electron-rich sample that
%shows a transition to a superconducting state below $T\approx$1.5K (sample
%thickness=1300~\AA ). (b) $\sigma(T)$ for In$_{x}$O sample with a much lower
%carrier concentration (but less disorder) exhibiting normal transport
%properties down to 0.3K (sample thickness =1000~\AA ). The dashed line is a
%fit to $\sigma(T)=\sigma_{0}+A\cdot T^{0.3}.$ The small value of the exponent
%in this case may suggest that the film is not far from crossing-over to a 2D
%behavior at the lower temperatures (which may result in a logarithmic
%dependence).}}{}{fig_5.eps}{\special{ language "Scientific Word";
%type "GRAPHIC";  maintain-aspect-ratio TRUE;  display "PICT";
%valid_file "F";  width 3.5405in;  height 2.9482in;  depth 0pt;
%original-width 9.5008in;  original-height 7.8975in;  cropleft "0";
%filename 'fig_5.eps';file-properties "XNPEU";}}}%
%BeginExpansion
\begin{figure}
[ptb]
\begin{center}
\includegraphics[
height=2.9482in,
width=3.5405in
]%
{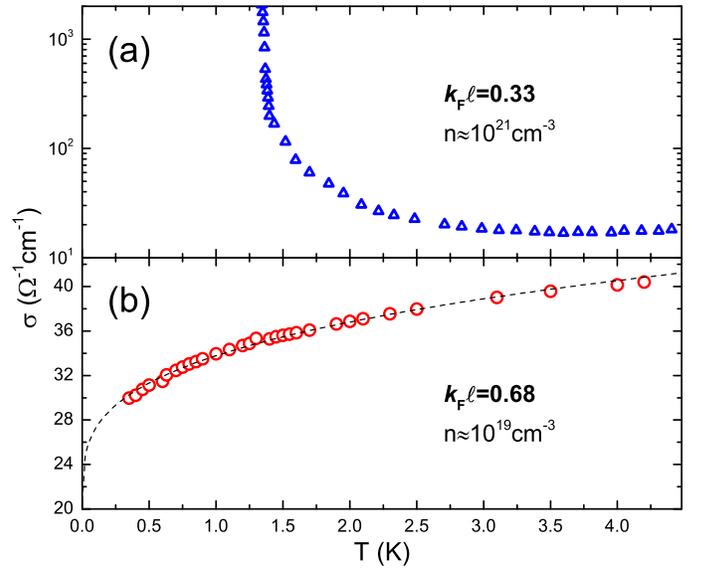}%
\caption{Conductivity as function of temperature plots for two In$_{x}$O
films; (a) Electron-rich sample that shows a transition to a superconducting
state below $T\approx$1.5K (sample thickness=1300~\AA ). (b) $\sigma(T)$ for
In$_{x}$O sample with a much lower carrier concentration (but less disorder)
exhibiting normal transport properties down to 0.3K (sample thickness
=1000~\AA ). The dashed line is a fit to $\sigma(T)=\sigma_{0}+A\cdot
T^{0.3}.$ The small value of the exponent in this case may suggest that the
film is not far from crossing-over to a 2D behavior at the lower temperatures
(which may result in a logarithmic dependence).}%
\end{center}
\end{figure}
%EndExpansion

The important corollary that follows from this observation is this: $k_{F}%
\ell$ \textit{is not a good parameter to characterize disorder when
superconductivity is concerned}. Reducing $k_{F}$ (by e.g., decreasing the
carrier concentration), weakens the BCS potential and the superconducting
transition temperature will decrease more than by reducing $\ell$ to achieve
the same $k_{F}\ell$. Increasing the static disorder will reduce $\ell$ but
this is not necessarily detrimental to superconductivity. It may actually
enhance it \cite{16}. Even in the case that superconductivity is suppressed by
disorder it is a weaker effect than the exponential decrease associated with
reducing the BCS potential; the latter can easily push the transition
temperature to well below experimental reach creating the false appearance of
a superconductor to \textit{metal} transition. $k_{F}\ell$ may still be a
descriptive measure of disorder when care is taken to keep $k_{F}$ constant in
the process of varying the disorder as was done in \cite{7}.

Despite the huge difference in their carrier concentration (more than two
orders of magnitude), the MIT in the two In$_{x}$O versions occurs at
essentially the same $k_{F}\ell$ (see Fig.4). Our conjecture, that the
apparent violation of the Ioffe-Regel criterion in the electron-rich In$_{x}$O
is due to correlations, is not supported. The low value of $k_{F}\ell$ at the
transition is therefore more likely related to some properties of the
amorphous phase that differ from the crystalline version of the material where
the critical $k_{F}\ell$ is close to unity.

The crystalline and amorphous phases of indium-oxide differ in several aspects
(in addition to symmetry). While the crystalline version, in agreement with
theoretical considerations \cite{17}, behaves as a nearly free electron system
\cite{18}, optical absorption due to interband transition of In$_{x}$O samples
are inconsistent with a parabolic conduction band \cite{6}. Values of
$k_{F}\ell$ estimated for this material by using free-electron formulae may
therefore be questionable. Another possibility is that transport in the system
is more inhomogeneous than might be judged by the space-filling, physically
uniform structure reported for this material \cite{7,19,20}. Note that when
the sample resistivity is not uniform, $k_{F}\ell$ based on macroscopic
measurements of conductance and carrier concentration may not be telling of
the relevant disorder and the calculated value of $k_{F}\ell$ may differ from
that associated with the `average' disorder. This could be a real problem when
the inhomogeneity in the system is so gross as to cause current to flow
preferentially through only part of the structure.

In the following we describe how some structural attributes of the amorphous
indium-oxides give rise to inhomogeneities in these systems (and possibly in
other multi-component systems like alloys and metal-oxides). These may lead to
a host of low temperature transport anomalies when superconductivity is
involved in addition to an underestimated $k_{F}\ell.$

The stoichiometric compound In$_{2}$O$_{3}$ is an ionic insulator with a large
($\approx$3.6eV) band gap. It is a well characterized material, has a cubic
structure with 48 oxygen atoms and 32 indium atoms in a unit cell. The
naturally occurring material however is oxygen deficient, and films of
crystalline indium-oxide usually contain 5-10\% oxygen vacancies \cite{21}.
This gives rise to carrier-concentration n$\approx$6$\cdot$10$^{\text{19}}%
$cm$^{\text{-3}},$ and a factor of $\pm$2 around this value may be affected by
changing the stoichiometry using, e.g., UV-treatment \cite{22}. The limited
range of achievable n is due to constraints imposed by crystal chemistry.

Amorphous indium-oxide, being relatively free of these constraints, may be
prepared with a much wider atomic ratio of oxygen-indium. This makes it
possible to make stable (more accurately; \textit{metastable }with a long
lifetime at room temperatures and below) films with optical gaps between 2.5eV
to 1.1eV and carrier concentration of $\approx$5$\cdot$10$^{\text{18}}%
$-10$^{\text{22}}$ cm$^{\text{-3}}$ respectively \cite{6}. The correlation
between stoichiometry and carrier concentration of amorphous indium-oxide is
shown in Fig.~6.%
%TCIMACRO{\FRAME{ftbpFU}{3.5267in}{2.9136in}{0pt}{\Qcb{Dependence of the
%carrier-concentration n on the In/O atomic ratio. The Carrier-concentration is
%based on room-temperature Hall Effect measurements and the O/In ratio was
%measured using Rutherford Backscattering (see ref. \cite{21} for technical
%details).}}{}{fig_6.eps}{\special{ language "Scientific Word";
%type "GRAPHIC";  maintain-aspect-ratio TRUE;  display "PICT";
%valid_file "F";  width 3.5267in;  height 2.9136in;  depth 0pt;
%original-width 9.8779in;  original-height 8.1206in;  cropleft "0";
%filename 'fig_6.eps';file-properties "XNPEU";}}}%
%BeginExpansion
\begin{figure}
[ptb]
\begin{center}
\includegraphics[
height=2.9136in,
width=3.5267in
]%
{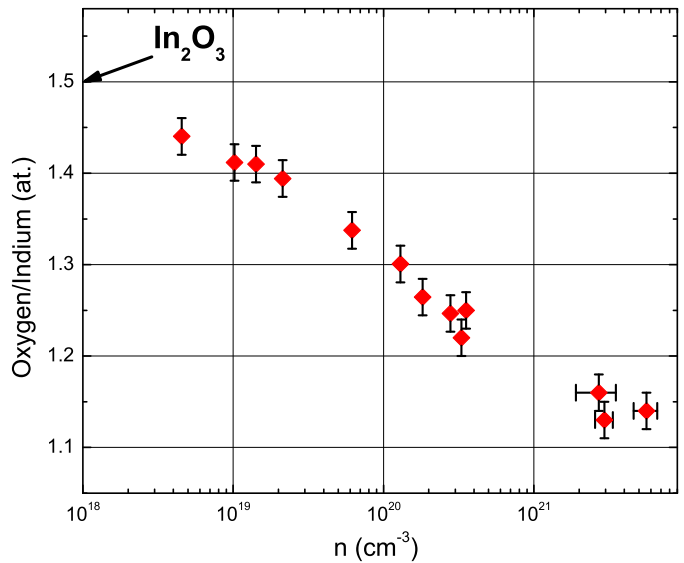}%
\caption{Dependence of the carrier-concentration n on the In/O atomic ratio.
The Carrier-concentration is based on room-temperature Hall Effect
measurements and the O/In ratio was measured using Rutherford Backscattering
(see ref. \cite{21} for technical details).}%
\end{center}
\end{figure}
%EndExpansion

The freedom from crystal chemistry constraints is perhaps also the reason for
the emergence of \textit{compositional}-disorder in these amorphous oxides.
This kind of spatial disorder has far reaching consequences, especially when
superconductivity is involved.

Spatial disorder means that the material parameters vary in space,
\textit{disordered }systems are\textit{ inhomogeneous }by definition.

The nature of the inhomogeneity however, depends on the \textit{type} of
disorder. Quenched disorder of the ionic potential is common in most
disordered electronic systems. Large spatial fluctuations of carrier
concentration on the other hand are unlikely to occur in systems with
mono-atomic metallic systems but they are quite prominent in all versions of
the amorphous indium-oxides. Figure 7 is a micrograph taken by a scanning mode
of a transmission electron microscope. The contrast mechanism in the
micrograph is due to absorption; changing the angle between the sample plane
and the electron beam axis does not turn a black region into white (as it
often would in a crystalline sample where the main contrast mechanism is Bragg
scattering). There are no holes in the film and thickness variations are
relatively small; AFM line scans show 5-20 \AA ~surface roughness for
200~\AA ~films deposited on glass substrates (the smaller value obtained for
the low-n version of the material). Chemical analysis, using energy dispersive
x-ray spectroscopy and electron energy-loss spectroscopy (EELS) revealed that
regions with higher transmission (white patches in the figure) are richer in
oxygen, poorer in indium content, and vice-versa for the black regions.
Variations in the O-In ratio between these regions could be as high as 15-40\%
for a sampling area of (50~\AA )$^{\text{2}}$. The uncertainty in these
measurements is mainly due to the background signal from the carbon support
that includes a certain amount of oxygen. This source of error may be
mitigated by making self-supporting films but a systematic study of thickness
dependence will be necessary to get a more accurate estimate of stoichiometry.
Note however, that a mere 10\% variation in O-In ratio is equivalent to an
\textit{order of magnitude} difference in the local carrier-concentration
(c.f., Fig.~6), which is tantamount to a factor of two in the thermodynamic
density of states.
%TCIMACRO{\FRAME{ftbpFU}{3.5405in}{2.22in}{0pt}{\Qcb{STEM (Scanning
%Transmission Electron Microscopy) picture of In$_{x}$O films. These are
%typical for the low-n version with carrier-concentration n=5$\cdot
%$10$^{\text{18}}$cm$^{\text{-3}}$ (a), and the electron-rich version with
%n=3$\cdot$10$^{\text{21}}$cm$^{\text{-3}}$ (b). The films are 300\AA ~thick
%(rather than the 1000-1400\AA ~thick films used for transport) to achieve a
%good spatial resolution and they were thermally annealed to make them
%characteristic of samples near their respective MIT (a) and SIT (b)
%transition.}}{}{fig_7.eps}{\special{ language "Scientific Word";
%type "GRAPHIC";  maintain-aspect-ratio TRUE;  display "PICT";
%valid_file "F";  width 3.5405in;  height 2.22in;  depth 0pt;
%original-width 8.3359in;  original-height 5.2001in;  cropleft "0";
%filename 'fig_7.eps';file-properties "XNPEU";}}}%
%BeginExpansion
\begin{figure}
[ptb]
\begin{center}
\includegraphics[
height=2.22in,
width=3.5405in
]%
{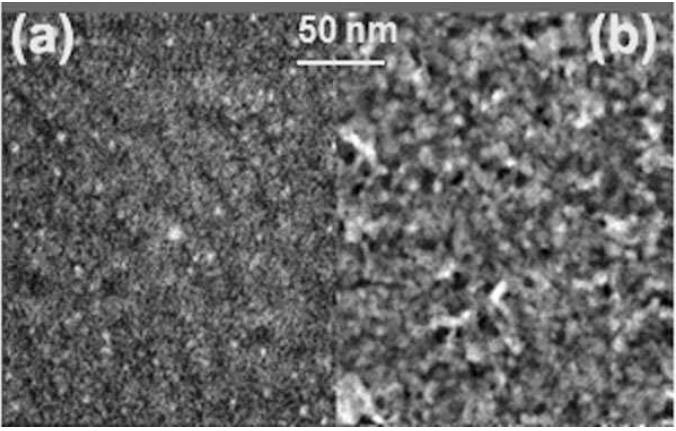}%
\caption{STEM (Scanning Transmission Electron Microscopy) picture of In$_{x}$O
films. These are typical for the low-n version with carrier-concentration
n=5$\cdot$10$^{\text{18}}$cm$^{\text{-3}}$ (a), and the electron-rich version
with n=3$\cdot$10$^{\text{21}}$cm$^{\text{-3}}$ (b). The films are
300\AA ~thick (rather than the 1000-1400\AA ~thick films used for transport)
to achieve a good spatial resolution and they were thermally annealed to make
them characteristic of samples near their respective MIT (a) and SIT (b)
transition.}%
\end{center}
\end{figure}
%EndExpansion

The spatial range of these compositional fluctuations can be assessed by
Fourier transforming line scans of the intensity, averaged over the entire
micrograph area (only a part of which is shown in Fig.~7). Results of such
analyses, done for the two In$_{x}$O samples of Fig.~7, are shown in Fig. 8.
The flattening out of the spectrum at the small spatial scales is partly due
to the smoothing effect of oxygen-diffusion and partly due to the STEM
resolution. Note that the low-n In$_{x}$O is somewhat more uniform than the
high-n version. However, in both cases the spectrum is skewed in a similar way
and the compositional fluctuations persist over scales extending up to
300-800~\AA .%

%TCIMACRO{\FRAME{ftbpFU}{3.5379in}{2.7397in}{0pt}{\Qcb{Fourier transforms of
%the STEM micrographs taken from the same samples shown in Fig.7 and plotted as
%function of the spatial scale. These were averaged over individual line scans
%taken across each micrograph. The dashed lines are merely guides to the eye.}%
%}{}{fig_8.eps}{\special{ language "Scientific Word";  type "GRAPHIC";
%maintain-aspect-ratio TRUE;  display "PICT";  valid_file "F";
%width 3.5379in;  height 2.7397in;  depth 0pt;  original-width 9.5415in;
%original-height 7.3812in;  cropleft "0";  croptop "1";  cropright "1";
%cropbottom "0";  filename 'fig_8.eps';file-properties "XNPEU";}}}%
%BeginExpansion
\begin{figure}
[ptb]
\begin{center}
\includegraphics[
height=2.7397in,
width=3.5379in
]%
{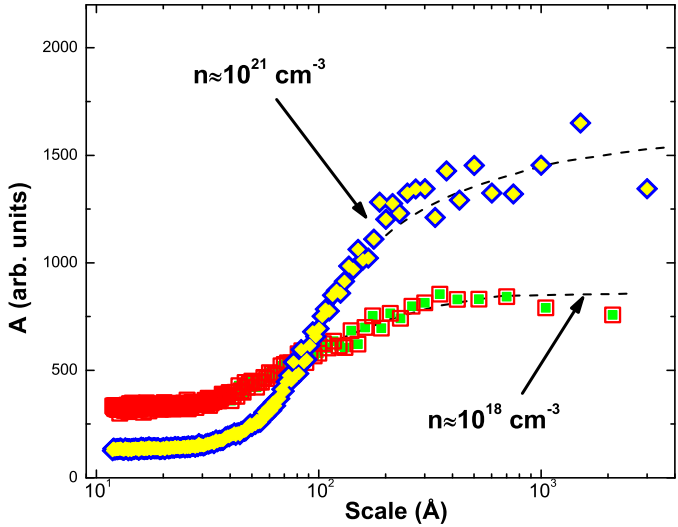}%
\caption{Fourier transforms of the STEM micrographs taken from the same
samples shown in Fig.7 and plotted as function of the spatial scale. These
were averaged over individual line scans taken across each micrograph. The
dashed lines are merely guides to the eye.}%
\end{center}
\end{figure}
%EndExpansion

These scales are comparable or even larger than the length scales that may be
relevant for transport; $L_{in},L_{T},\ell,$ (and $\xi_{s}$, the
superconducting coherence-length). Pertinent information on these transport
parameters may be obtained by analyzing magneto-conductance (MC) data. We have
measured ten In$_{x}$O metallic samples at 4K, and for comparison, also at
77K. At small fields the MC for all metallic samples was positive and that
remained so down to 2K (tested on one of these samples). Spin orbit scattering
in In$_{x}$O is therefore rather limited in strength. A negative MC component
however does appear at high fields as seen in the data described in figure 9.

Figure 9 compares the MC of our most metallic In$_{x}$O sample with a In$_{2}%
$O$_{3-x}$ sample of comparable resistivity. The latter system has been
extensively studied and exhibits $L_{in}$ of 1000-1300~\AA ~at $\approx$4K for
samples with $k_{F}\ell$ of order 2-5 \cite{23}. The field at which the MC
crosses over from H$^{\text{2}}$ to a weaker dependence is $\approx$25 times
larger for the In$_{x}$O sample. This suggests a considerably smaller $L_{in}$
than that of In$_{2}$O$_{3-x}$ at the same temperature. We have analyzed the
low field MC for the amorphous sample using Kawabata's \cite{24} expression:
$\frac{\Delta\sigma}{\sigma}$=$\frac{\tau_{el}^{\text{1/2}}\tau_{in}%
^{\text{3/2}}e^{\text{2}}H^{\text{2}}}{12\sqrt{3}m^{\text{2}}}$ in conjunction
with the Drude expression for the conductivity $\sigma$=$\frac{ne^{\text{2}}%
}{m^{\ast}}\tau_{el}$ using m*$\approx$0.28$\cdot$(mass of the free electron).
This yielded the parameters: $\tau_{el}\approx$6$\cdot$10$^{\text{-15}}~$s,
$\ell\approx$12~\AA ~(using $V_{F}\approx$2$\cdot$10$^{\text{7}}~$cm/s),
$L_{in}$(at 4.1K)=450$\pm$20~\AA , $D\approx$2~cm$^{\text{2}}$/s. The
interaction length is then 70~\AA ~and 140~\AA ~depending on whether $L_{T}%
$=$k_{F}^{-1}(\frac{T}{T_{F}})^{\frac{1}{\eta}}$ or $\sqrt{\frac{\hbar
D}{k_{B}T}}$ respectively. At temperatures relevant for our experiments these
length scales are not large enough to average out the compositional
inhomogeneities observed in these materials.%
%TCIMACRO{\FRAME{ftbpFU}{3.5129in}{2.9066in}{0pt}{\Qcb{Magneto-conductance for
%a metallic In$_{x}$O film ($k_{F}\ell$=0.68) shown for 77K and 4.11K. These
%are compared with the magneto-conductance for a crystalline indium-oxide
%sample (with $k_{F}\ell$=2.3) taken at 4.11K. Dashed lines depict the
%$\Delta\sigma/\sigma\propto H^{2}$ law expected for small fields. Arrows mark
%the fields where $\Delta\sigma/\sigma$ deviates from the parabolic law. The
%ratio $\Delta\sigma/\sigma$(4K)$/\Delta\sigma/\sigma$(77K) measured for ten
%In$_{x}$O samples in the parabolic part of the MC was 78$\pm$4. Using
%Kawabata's expression \cite{24} $\Delta\sigma/\sigma\propto\tau_{in}^{3/2}$
%this is consistent with $\tau_{in}\propto T^{-1}$ in this range of
%temperatures. Note the negative MC of the In$_{x}$O sample (at 4.1K) for
%H\TEXTsymbol{>}6.5T.}}{}{fig_9.eps}{\special{ language "Scientific Word";
%type "GRAPHIC";  maintain-aspect-ratio TRUE;  display "USEDEF";
%valid_file "F";  width 3.5129in;  height 2.9066in;  depth 0pt;
%original-width 9.5415in;  original-height 7.3812in;  cropleft "0";
%croptop "1";  cropright "0.9920";  cropbottom "0";
%filename 'fig_9.eps';file-properties "XNPEU";}}}%
%BeginExpansion
\begin{figure}
[ptb]
\begin{center}
\includegraphics[
trim=0.000000in 0.000000in 0.076332in 0.000000in,
height=2.9066in,
width=3.5129in
]%
{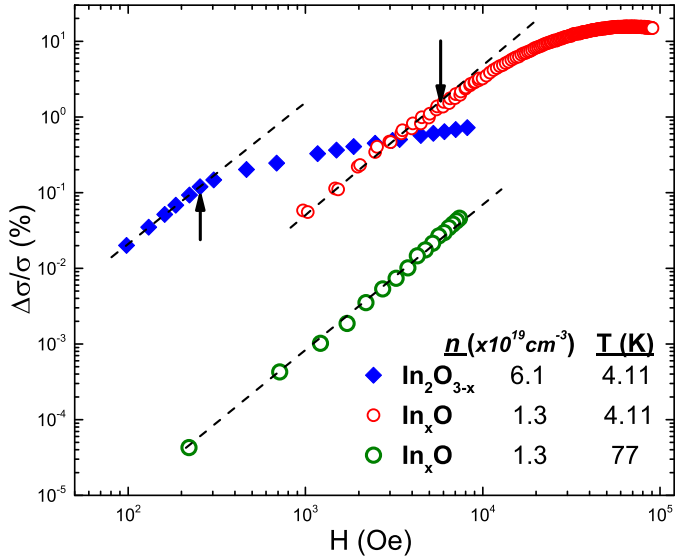}%
\caption{Magneto-conductance for a metallic In$_{x}$O film ($k_{F}\ell$=0.68)
shown for 77K and 4.11K. These are compared with the magneto-conductance for a
crystalline indium-oxide sample (with $k_{F}\ell$=2.3) taken at 4.11K. Dashed
lines depict the $\Delta\sigma/\sigma\propto H^{2}$ law expected for small
fields. Arrows mark the fields where $\Delta\sigma/\sigma$ deviates from the
parabolic law. The ratio $\Delta\sigma/\sigma$(4K)$/\Delta\sigma/\sigma$(77K)
measured for ten In$_{x}$O samples in the parabolic part of the MC was 78$\pm
$4. Using Kawabata's expression \cite{24} $\Delta\sigma/\sigma\propto\tau
_{in}^{3/2}$ this is consistent with $\tau_{in}\propto T^{-1}$ in this range
of temperatures. Note the negative MC of the In$_{x}$O sample (at 4.1K) for
H$>$6.5T.}%
\end{center}
\end{figure}
%EndExpansion

Although it may have no direct bearing on the main issues of this work, it is
important to point out that the MC of the In$_{x}$O samples differs from the
MC in the crystalline version by more than the magnitude of $L_{in}$. The
negative MC component of the metallic sample in Fig.~9, that is just
observable above $\approx$6.5T, becomes quite noticeable at lower temperatures
(but down to 1.4K it is only seen above $\approx$3T). The MC in diffusive
In$_{2}$O$_{3-x}$ samples is purely positive, throughout the entire range up
to fields of 12T and down in temperature from 77K to $\approx$40mK. There is
also a difference between In$_{x}$O and the crystalline In$_{2}$O$_{3-x}$
version in terms of the MC for insulating samples; The MC in insulating
In$_{2}$O$_{3-x}$ starts positive and becomes negative at high fields
\cite{25} while just the opposite is observed in our insulating In$_{x}$O
sample (with $k_{F}\ell$=0.23) for temperatures below 3K.

Results of MC measurements on In$_{x}$O films were reported by Lee et al
\cite{26}. Their results differed from ours; they observed positive MC only
above 8-10K while the low field MC became negative at lower temperatures. It
should be noted however that the In$_{x}$O films of Lee et al were presumably
the electron-rich version (and exhibited superconductivity for $T$%
%TCIMACRO{\TEXTsymbol{<}}%
%BeginExpansion
$<$%
%EndExpansion
1K when the resistance was low enough). More work is needed to identify the
origin of the mechanism that is responsible for the negative MC at both the
metallic and insulating regime of the different versions of In$_{x}$O.

Fluctuations in carrier concentrations on the spatial scales observed in the
amorphous indium-oxides constitute an important type of disorder. Screening is
more effective in regions with larger n, which means that the conductance in
these regions is higher (although the mobility may be impaired in these
regions, partially offsetting this effect). When the sample resistivity is not
uniform, $k_{F}\ell$ based on macroscopic measurements of conductance and
carrier concentration will in general be underestimated; since current
naturally flows preferentially through the more conducting regions of the
structure, the "active" volume for transport is smaller than that assumed by
its geometric dimensions. The non-uniform current distribution affects the
perceived values of both, the conductivity $\sigma,$ and the
carrier-concentration $n$ which are used in assigning a value of $k_{F}\ell$
to a given sample. This is based on $k_{F}\ell$=(3$\pi^{\text{2}}%
$)$^{\text{2/3}}\frac{\hbar\sigma_{\text{RT}}}{e^{\text{2}}n^{\text{1/3}}},$
and therefore errors in $n$ have a lesser effect than errors in $\sigma
_{\text{RT}}$ unless the system inhomogeneity is excessive (the Hall voltage
may be significantly suppressed in e.g., filamentary conduction yielding a
spuriously huge $n$). It seems intuitively plausible that, in general,
inhomogeneity causes $\sigma_{\text{RT}}$ to be underestimated while the
Hall-derived $n$ is if anything somewhat overestimated. These consideration
may explain why the metal to insulator transition in the two versions of
In$_{x}$O occurs for $k_{F}\ell$ that is smaller than unity. At the same time,
for scales of the order of millimeters, these compositional inhomogeneities
are well-averaged, and therefore the excellent uniformity of the In$_{x}$O
films mentioned in section II is not at variance with the observed mesoscale inhomogeneity.

A more profound modification of transport should occur at low temperatures
when superconductivity appears. This is where the value of $k_{F}\ell$ as a
parameter that characterizes disorder has questionable merit; carrier-rich
regions of the system are more likely to go superconducting than regions where
n is small. As demonstrated in Fig.~5, the sample with the larger carrier
concentration goes superconducting despite having a \textit{smaller}
$k_{F}\ell$ than a sample with a larger mean-free-path. It is the BCS
potential that plays the crucial role in these situations, and when its local
value fluctuates in the system on scales that are comparable or larger than
$\xi_{s}$, inhomogeneity is dramatically enhanced. In this case the system
breaks into a `granular-like' structure; islands of zero resistance
precipitate in a normal matrix. The latter may be either metallic or
insulating depending on the (average) degree of disorder, carrier
concentration, and temperature. Due to proximity and Josephson effects (that
are temperature dependent), the scales of the superconducting islands and
their spatial distribution may somewhat differ from those set by the
compositional disorder.

The following transport anomalies found in the electron-rich version of
In$_{x}$O are in line with this picture:

1) The\ critical disorder for the SIT is scale dependent; short samples may go
superconducting while longer ones, made on the same strip and having
essentially the \textit{same} (normal state) resistivity, exhibit
\textit{insulating} behavior \cite{20}.

2) On the just insulating side of the transition, where the conductance versus
temperature follows $\sigma(T)\propto$exp$(-T_{0}/T)$, the activation energy
$T_{0}$ increases monotonously with the sample length \cite{20}.

3) The current voltage characteristics of these samples revealed discontinuous
jumps similar to those found in Josephson arrays \cite{19}.

4) A non-monotonous MC is observable in In$_{x}$O films near their SIT; The
conductance first decreases with field reaching, in some cases, a much smaller
value then its asymptotic value at high fields \cite{27}.

Significantly, all these anomalies appear at the temperature range where the
less disordered material is superconducting.

Other possible manifestations of superconducting islands coexisting with
normal regions were obtained in \cite{28} and in recent tunneling experiments
from large area electrodes into In$_{x}$O films in the insulating regime
\cite{29}. It may be difficult to observe these non-superconducting regions by
a scanning tunneling technique (STM) due to the spreading resistance
associated with resistive paths but the inhomogeneous nature of
superconductivity has been seen by STM in In$_{x}$O films \cite{30} (and in
TiN \cite{31}).

The observation that the In$_{x}$O films are structurally continuous, and
being amorphous, free of grain boundaries etc., was an inducement to offer an
inherent `disorder induced granularity' scenario for these phenomena
\cite{19,20}. The effects associated with this scenario ought to have been
apparent in \textit{all} materials near their SIT provided that
superconductivity is not suppressed by disorder before a `critical
$G$'$\approx\frac{e^{\text{2}}}{h}$ is reached \cite{19,20}. Several systems
such as TiN \cite{32} and NbN \cite{33} indeed exhibited some of these
transport anomalies near their SIT (specifically, a non-monotonous MC).
However, the magnitude of these effects were not as prominent as they are in
the electron-rich In$_{x}$O samples. Moreover, these peculiarities are not
usually observed near the \textit{disorder-tuned} \cite{34} SIT of amorphous
Bi and Pb films, except when structural granularity was deliberately
introduced to modify their microstructure \cite{35}.

We have therefore to conclude that, while the `inherent disorder-induced'
inhomogeneity is a plausible physical scenario, it appears that the
modulation-depth of local superconducting properties caused by potential
fluctuations is much less conspicuous than that of fluctuations in
carrier-density. In other words, the \textit{type} of disorder is an important
ingredient in producing prominent `granularity effects' near the SIT. Unless
they break time-reversal symmetry, potential fluctuation have a mild effect on
$T_{c}$ (which can be of either sign \cite{16}). Spatial fluctuation in the
BCS potential, on the other hand, is a powerful agent for local modulation of
$T_{c}$, and it may facilitate the two-phase state responsible to the observed
transport peculiarities. An effective way to achieve it is by modulating
carrier concentration, which seem to occur naturally in the amorphous indium-oxides.

This type of disorder is energetically unfavorable in a mono-atomic system
like amorphous Bi unless the system is physically discontinuous. On the other
hand, in an alloy or metal oxide, compositional-disorder may compensate for
the carrier concentration difference by a chemical one without compromising
the physical continuity of the structure. Grain boundaries and other extended
defects may also play a role in creating inhomogeneous BCS potential
distribution when pair-breakers such as magnetic impurities (that tend to
segregate at these defects) are present in the system.

The corollary that emerges from these considerations is that while disorder is
tantamount to inhomogeneity, the degree and detailed nature of the resulting
spatial fluctuation depend on the \textit{type }of disorder, not just on its
magnitude. This distinction may be less important for the non-interacting
system but it is crucial for the subtle case of superconductivity.

One of the questions raised in the study of insulating In$_{x}$O films near
the SIT \cite{7} was the origin of the dependence of the conductance on
temperature $\sigma(T)\propto$exp$(-T_{0}/T)$. An arrhenius law for
$\sigma(T)$ is intriguing in that it occurs in a disordered system for which
the natural choice is the variable-range-hopping (VRH) mechanism \cite{5},
which indeed is exhibited by just-insulating samples of the low-n version of
In$_{x}$O (Fig.~3b). The study reported in \cite{7} used a three-dimensional
system. The activation energy of the insulating samples increased monotonously
with disorder reaching a mximum of $T_{0}\approx$7K before the conduction
mechanism reverted to variable range hopping \cite{7}. It was shown in a
subsequent study using a \textit{two}-dimensional version of this system that
$T_{0}$ may reach 14-15K \cite{19} for a similar range of $k_{F}\ell$ used in
the 3D system. The dimensionality and length dependence of $T_{0}$ are
characteristic signs of a percolative phenomenon.

These observations suggest that the origin of the $\sigma(T)\propto
$exp$(-T_{0}/T),$ or a somewhat \textit{faster }dependence \cite{19}, is the
appearance of superconducting islands in an insulating medium. In the
insulating regime of the SIT (and at temperatures where the material is fully
in the normal state), most of the electron-rich areas are weakly coupled to
other regions, say with typical coefficient $t\ll$1. As the temperature falls
below the local $T_{c}$ of these regions, the coupling (being now controlled
by Andreev processes), becomes $t^{2}\lll$1. Superconducting islands that were
part of the current-carrying-network (CCN) in the normal state, will be
effectively removed from it thereby forcing a new and less conductive CCN.
Transport in the VRH regime takes place in a tenuous current-carrying-network
that, for a given disorder, becomes more rarefied as temperature is lowered.
An \textit{exponential} reduction of the conductance occurs when part of the
sample volume is eliminated from the transport \cite{36}, a role that the
superconducting islands may effectively take.

The range of disorder where this mechanism is relevant is limited to the
vicinity of the transition; For sufficiently strong disorder the localization
length $\xi$ will be everywhere smaller than $\xi_{s}$ and there will be no
superconducting islands in the system. In this case the transport mechanism
will revert to normal VRH as seen in the experiment \cite{7}. At small
disorder, on the other hand, many of the superconducting islands may be
Josephson-coupled to form large superconducting clusters. Effective Josephson
coupling through \textit{insulating} In$_{x}$O layers has been observed
experimentally \cite{37}. Global superconductivity may result in the system
with the associated percolative features discussed in reference \cite{20}. In
this picture $\sigma(T)$ is a non-trivial result of hopping conductivity
modulated by the compounded effect of the temperature-dependent appearance of
superconducting islands \textit{and} the evolution with temperature of their
local pair-potential. In view of the transport features described above this
seems to be a prime avenue to look for a solution to this problem.

This work greatly benefitted from an extensive exchange of data and ideas with
the research groups of Allen Goldman, and Aviad Frydman. We also acknowledge
illuminating discussion with O. Agam, D. Khmel'nitskii, Y. Imry, and M.
Feigel'man. One of us (Z.O.) expresses his gratitude to F. P. Milliken for
help in getting some of the Rutherford Backscattering data at the IBM research
Center, Yorktown-Heights. This research has been supported by the Binational
US-Israel Science Foundation and by The Israeli Academy for Sciences and Humanities.


\begin{thebibliography}{99}
\bibitem {13}Samples with $k_{F}\ell\gtrsim$0.25 in this electron-rich version
of In$_{x}$O are superconducting at liquid-helium temperatures and therefore
the critical $k_{F}\ell$ cannot be determined by the low-temperature method
used for the non-superconducting In$_{2}$O$_{3-x}$and the low-n version of
In$_{x}$O.
\end{thebibliography}
\end{document}